\documentclass[12pt]{article}
\usepackage{amsfonts}
\usepackage{amsmath}
\usepackage[top=2cm, bottom=1.5cm, left=2cm, right=2cm]{geometry}
\usepackage[colorlinks=true, linkcolor=red, citecolor=blue]{hyperref}

\makeatletter

\@addtoreset{equation}{section}
\makeatother

\usepackage{float}
\usepackage{mathtools}
\usepackage{graphicx}
\usepackage[font=small,labelfont=bf,labelsep=space]{caption}
\usepackage{subfig}

\input{tcilatex}
\begin{document}

\title{Coherent states associated with tridiagonal Hamiltonians}
\author{Hashim A. Yamani${}^{\ast }$ and Zouha\"{\i}r Mouayn$^{\natural }$} 
\maketitle
\vspace*{-1em}
\begin{center}
{\footnotesize ${}^{\ast }$  Dar Al-Jewar, Knowledge Economic City, Medina, Saudi Arabia\vspace*{-0.2em}\\ 
e-mail: hashim.haydara@gmail.com
\vspace*{0.6mm}\\[3pt]
${}^{\natural  }$ Department of Mathematics, Faculty of Sciences
and Technics (M'Ghila),\vspace*{-0.2em}\\ Sultan Moulay Slimane University, B\'{e}ni Mellal, Morocco \vspace*{-0.5em}\\ e-mail: mouayn@usms.ma
 }
\end{center}

\begin{abstract}
It has been shown that a positive semi-definite Hamiltonian $H$, that has a
tridiagonal matrix representation in a given basis, can be represented in
the form $H=A^{\dag }A$, where $A$ is a forward shift operator playing the
role of an annihilation operator. Such representation endows $H$ with rich
supersymmetric properties yielding results analogous to those obtained by
studying the Hamiltonian as a differential operator. Here, we study the
coherent states which we define as being the eigenstates of the operator $A$%
. We explicitly find the expansion coefficients of these states in the given
basis. We further identify a complete set of special coherent states which
themselves can be used as basis. In terms of these special coherent states,
we show that a general coherent state has the expansion form of a Lagrange
interpolation scheme. As application of the developed formalism, we work out
examples of systems having pure discrete, pure continuous, or mixed energy
spectrum.
\end{abstract}

\section{Introduction}

Coherent states (CS) have been introduced by Schr\"odinger as states which
behave in many respects like classical states \cite{Sch}. They acquired this name
when Glauber \cite{G} realized that they were particularly convenient
to describe optical coherence. In particular, the electromagnetic radiation
generated by a classical current is a multimode coherent state. So is
the light produced by a laser in certain regimes \cite{Schl}. Thereafter, CS
became cornerstones of modern quantum optics \cite{KS} and more recently, CS
found applications in quantum information experiments \cite{GVWB}.

CS also are mathematical tools which provide a close connection between
classical and quantum formalisms so as to  play a central role in the
semiclassical analysis \cite{AAG, JPG}. In general, CS are 
overcomplete family of normalized \textit{ket} vectors $\left\vert \zeta
\right\rangle $ which are labeled by points $\zeta $ of a phase-space domain 
$X$ , blonging to a Hilbert space $\mathcal{H}$ that corresponds to a
specific quantum model and provide $\mathcal{H}$ with a resolution of its
identity operator as 
\begin{equation}\label{eq1.1}
\mathbf{1}_{\mathcal{H}}=\int\limits_{X}\left\vert \zeta \right\rangle
\left\langle \zeta \right\vert d\mu \left( \zeta \right) .  
\end{equation}
with respect to a suitable integration measure $d\mu \left( \zeta \right) $
on $X.$ These CS are constructed for that space with $\mathcal{H}$ having either a
discrete or continuous basis in different ways \cite{Do} : ``
\textit{\`{a} la Glauber}'' as eigenfunctions of an annihilation operator; as
states minimizing some uncertainty principle or they can be obtained ``
\textit{\`{a} la} \textit{Gilmore-Perelomov}'' as orbits of a unitary
operator acting on a specific or fiducial state by appealing to the
representation theory of Lie groups \cite{Pe1} . Their number
states expansion over the eigenstates basis of the Hamiltonian has also lead
to a generalization which is known as the \textit{Hilbertian} \textit{%
probabilistic }scheme\textit{\ }\cite{JPG}.

These states have long been known for the Hamiltonian of the harmonic
oscillator, whose properties have been used as a model for more quantum
Hamiltonian systems. Most of theoretical or mathematical research treats the
system's Hamiltonian as a differential operator. Here we examine its matrix
representation in a space spanned by complete square integrable basis
functions. In particular, we focus on Hamiltonians whose matrix
representations are of the tridiagonal form. This in not a restriction since
tridiagonality can always be achieved by a complete new basis constructed
from a linear combination of the original basis. In fact, using an algorithm
such as that of Lanczos \cite{Lan50, deh81}, one can construct such a basis
from only one seed function.

The advantage of tridiagonality is that it makes the coefficients of expansion of  the
eigenvector in the basis satisfy recursion relations similar to those satisfied by
orthogonal polynomials. In fact, the rich tools of the field of orthogonal
polynomials and the associated spectral theory have been brought to bear to
mathematically underpin the properties of tridiagonal Hamiltonians, making
this approach an active field of research. Having investigated the
supersymmetry of such Hamiltonians \cite{YM2}, we now investigate a closely related
topic of the coherent states associated with these Hamiltonians.

The paper is organized as follows. In section 2, we review the
needed  properties of tridiagonal Hamiltonians and define the
coherent states associated with them. In section 3, we explicitly solve for
the coefficients of the coherent state vector expansion in the basis, and
then clarify how to compute the time development of the resulting state. In
section 4, we examine and characterize the properties of a special set of
coherent states. In particular, we show that the set is complete. This
property qualities the special set of coherent states to be used as a basis.
We then proceed to find the expansion coefficients of a general coherent
states in term of the special basis. We conclude by showing that the
resulting expansion is completely analogous to the known Lagrange
interpolation scheme. Borrowing from known results regarding this scheme, we
immediately write down several different but equivalent forms for such
expansion. In section 5, we work out several examples illustrating  the
results in the previous sections. We conclude in section 6 with some
comments.

\section{Review of the properties of tridiagonal Hamiltonians}

We assume that we are given a positive semi-definite Hamiltonian $H$ (with
zero as the value of the lowest energy in its spectrum) acting on a Hilbert
space $\mathcal{H}$, that has a tridiagonal matrix representation in the
orthonormal basis $\{{\left\vert \phi _{n}\right\rangle }\}_{n=0}^{\infty }$
of $\mathcal{H}$ with known coefficients $\{a_{n},b_{n}\}_{n=0}^{\infty }$ 
\begin{equation}
H_{n,m}={\left\langle \phi _{n}\right\vert }H{\left\vert \phi
_{m}\right\rangle }=b_{n-1}\,\delta _{n,m+1}+\,a_{n}\,\delta
_{n,m}\,+\,b_{n}\,\delta _{n,m-1}.  \label{GrindEQ__2.1_}
\end{equation}%
We solve the energy eigenvalue equation $H{\left\vert E\right\rangle }=E{%
\left\vert E\right\rangle }$ by expanding the eigenvector ${\left\vert
E\right\rangle }$ in the basis ${\left\vert \phi _{n}\right\rangle }$ as ${%
\left\vert E\right\rangle }=\sum_{n=0}^{\infty }f_{n}(E){\left\vert \phi
_{n}\right\rangle }$. Making use of the tridiagonality of $H$, we readily
obtain the following recurrence relations for the expansion coefficients 
\begin{equation}
\begin{array}{l}
{Ef_{0}(E)=a_{0}f_{0}(E)+b_{0}f_{1}(E)\quad \quad } \\ 
{Ef_{n}(E)=b_{n-1}f_{n-1}(E)+a_{n}f_{n}(E)+b_{n}f_{n+1}(E),\quad
n=1,2,...,.\quad }%
\end{array}
\label{GrindEQ__2_}
\end{equation}%
The operator $H$ may admit, in addition to the continuous spectrum $\sigma
_{c}$, a discrete part $\{E_{\mu }\}_{\mu }$ both of which lead to the
following form of the resolution of the identity operator on $\mathcal{H}$ 
\begin{equation}
\sum_{\mu }{\left\vert E_{\mu }\right\rangle }{\left\langle E_{\mu
}\right\vert }+\int_{\sigma _{c}}{\left\vert E\right\rangle }{\left\langle
E\right\vert }\,dE\;=1_{\mathcal{H}}.\quad \quad  \label{GrindEQ__3_}
\end{equation}%
This translates into the following orthogonality relation for the expansion
coefficients 
\begin{equation}
\sum_{\mu }f_{n}(E_{\mu })\left( f_{m}(E_{\mu })\right) ^{\ast
}+\int_{\sigma _{c}}f_{n}(E)\left( f_{m}(E)\right) ^{\ast }dE\,=\delta _{n,m}.
\label{GrindEQ4}
\end{equation}%
If we now define $p_{n}(E)=\frac{f_{n}(E)}{f_{0}(E)}$, then $\{p_{n}(E)\}$ is
a set of polynomials that satisfy the three-term recursion relation 
\begin{equation}
Ep_{n}(E)=b_{n-1}p_{n-1}(E)+a_{n}p_{n}(E)+b_{n}p_{n+1}(E),\quad n=1,2,...,
\label{GrindEQ__5_}
\end{equation}%
with the initial conditions $p_{0}(E)=1$ and $p_{1}(E)=(E-a_{0})b_{0}^{-1}$.
If we further define $\Omega (E)=\left\vert f_{0}(E)\right\vert ^{2}$and $%
\Omega _{\mu }=\left\vert f_{0}(E_{\mu })\right\vert ^{2}$, then the
relation \eqref{GrindEQ4} now translates into the following orthogonality
relation for the polynomial $p_{n}$ 
\begin{equation}
\sum_{\mu }\Omega _{\mu }p_{n}(E_{\mu })\left( p_{m}(E_{\mu })\right) ^{\ast
}+\int\limits_{\sigma _{c}}\Omega (E)p_{n}(E)\left( p_{m}\left( E\right)
\right) ^{\ast }dE\,=\delta _{n,m}.  \label{GrindEQ__6_}
\end{equation}%
We have shown \cite{YM2} that we can write the Hamiltonian $H$ in the form $%
H=A^{\dag }A$ where the forward-shift operator $H$ is defined by its action
on the basis vector as 
\begin{equation}
A{\left\vert \phi _{n}\right\rangle }=c_{n}\,{\left\vert \phi
_{n}\right\rangle }+\,d_{n}\,{\left\vert \phi _{n-1}\right\rangle }\,
\label{GrindEQ__7_}
\end{equation}%
where the coefficients $\{c_{n},d_{n}\}_{n=0}^{\infty }$ are related to the
coefficients $\{a_{n},b_{n}\}_{n=0}^{\infty }$ and the polynomials $%
\{p_{n}\}_{n=0}^{\infty }$ 
\begin{equation}
{d_{0}=0,\,\;c_{n}^{2}=-b_{n}\frac{p_{n+1}(0)}{p_{n}(0)}\quad ,\quad
d_{n+1}^{2}=-b_{n}\frac{p_{n}(0)}{p_{n+1}(0)},\quad \quad } {a_{n}=c_{n}^{2}+d_{n}^{2}\quad ,\quad b_{n}=c_{n}^{\ast }d_{n}}.%
\label{GrindEQ__8_}
\end{equation}%
We define the coherent states associated with the tridiagonal Hamiltonian as
the eigenstates of the operator $A$.

\section{General solution of the coherent state}

\subsection{General solution at $t=0$}

We write the coherent state~$\vert z)$ as the normalized
solution to the eigenvalue equation 
\begin{equation}
A\vert z) =z\vert z). \label{GrindEQ__9_}
\end{equation}%
We expand the state in terms of the basis as 
\begin{equation}
\vert z)=\sum_{n=0}^{\infty }\Lambda _{n}(z)\,{\left\vert
\phi _{n}\right\rangle }.  \label{GrindEQ__10_}
\end{equation}%
Using Eq.\eqref{GrindEQ__7_}, this equation gives 
\begin{equation}
\sum_{n=0}^{\infty }z\Lambda _{n}(z)\,{\left\vert \phi _{n}\right\rangle }%
=\sum_{n=0}^{\infty }\left[ c_{n}\Lambda _{n}(z)+\,d_{n+1}\Lambda _{n+1}(z)\right]%
\vert \phi _{n}\rangle.  \label{GrindEQ__11_}
\end{equation}%
This yields the recursion solution for the coefficients $\{\Lambda
_{n}(z)\}_{n=0}^{\infty }$ , namely, 
\begin{equation}
\Lambda _{0}(z)=\left( \sum_{n=0}^{\infty }\left\vert Q_{n}(z)\right\vert
^{2}\right) ^{-\frac{1}{2}}\text{and }\Lambda _{n}(z)=\Lambda
_{0}(z)\,Q_{n}(z),\text{ }n\geq 1  \label{GrindEQ__12_}
\end{equation}%
where%
\begin{equation}
Q_{0}(z)=1\text{ and}\;Q_{n}(z)=\prod_{j=0}^{n-1}(\frac{z-c_{j}}{d_{j+1}}%
),\;n\geq 1.\;
\end{equation}%
This general representation of the systems coherent state has two special
cases. The first one pertains to the fact that the ground state ${\left\vert
E=0\right\rangle }$ is itself a coherent state $\vert z)$
with $z=0$. To see this, we use the expansion of the ground state ${%
\left\vert E=0\right\rangle }=\sum_{n=0}^{\infty }f_{n}(0){\left\vert \phi
_{n}\right\rangle }$ . Hence,%
\begin{equation}
A{\left\vert E=0\right\rangle }=\sum_{n=0}^{\infty }\left[
c_{n}f_{n}(0)+d_{n+1}f_{n+1}(0)\right] {\left\vert \phi _{n}\right\rangle }.
\end{equation}
But we know from Eq.\eqref{GrindEQ__8_} that 
\begin{equation}
\frac{f_{n+1}(0)}{f_{n}(0)}=\frac{p_{n+1}(0)}{p_{n}(0)}=-\frac{c_{n}}{d_{n+1}%
}.
\end{equation}%
Thus, $A{\left\vert E=0\right\rangle }=0$. In section 5, we give two
examples of how to use the general form of the solution in Eq. %
\eqref{GrindEQ__10_} to find the explicit form of the ground states. The second special case is when the
Hamiltonian has a known pure discrete energy spectrum. The Hamiltonian, of
course, has a diagonal matrix representation in the basis composed of the
system's energy eigenvectors. Since diagonal matrices are special
tridiagonal ones, the above conditions leading to Eq. \eqref{GrindEQ__10_}
apply. In fact, all the $c_{n}$vanish and $d_{n}=\sqrt{E_{n}}$. Then,%
\begin{equation}
Q_{n}(z)=\frac{z^{n}}{\sqrt{(E_{n})!}}
\end{equation}
where $(E_{0})!=1$ and $(E_{n})!=E_{1}...E_{n}$ for $n\geq 1$ denotes the generalized factorial. This
simplifies the expansion sum Eq.\eqref{GrindEQ__10_} and may renders it
doable. Even more, because we know how the energy eigenstates develop in
time, each term contributing to sum in the expansion picks up a factor $%
e^{-iE_{n}t}$. In section 6 we illustrate this by considering the case of the one-dimensional harmonic oscillator.

\subsection{Time development of the coherent states}

If we know the details of the spectrum of the Hamiltonian, we can then
describe the time-development of the coherent state $\left\vert \left.
z\right) \right. $. We can invert the expansion of the energy eigenvector in
term of the basis function to write the basis function in term of the
complete energy eigenstates using the resolution of the identity Eq. %
\eqref{GrindEQ__3_} resulting in 
\begin{equation}
{\left\vert \phi _{n}\right\rangle }=\sum_{\mu }\sqrt{\Omega _{\mu }}%
p_{m}^{\ast }(E_{\mu }){\left\vert E_{\mu }\right\rangle }%
+\int\limits_{\sigma _{c}}\sqrt{\Omega (E)}p_{m}^{\ast }(E){\left\vert
E\right\rangle }dE.  \label{GrindEQ__13_}
\end{equation}%
Substituting back in the expansion of the coherent states, we obtain the
coherent eigenstate in term of the energy eigenstate of the Hamiltonian, 
\begin{equation}
\vert z)=\Lambda _{0}(z)\sum_{n=0}^{\infty }Q_{n}(z)\,%
\left[ \sum_{\mu }\sqrt{\Omega _{\mu }}p_{m}^{\ast }(E_{\mu }){\left\vert
E_{\mu }\right\rangle }+\int\limits_{\sigma _{c}}\sqrt{\Omega (E)}%
p_{m}^{\ast }(E){\left\vert E\right\rangle }dE\right].  \label{GrindEQ__14_}
\end{equation}%
Because we know the time-development of the energy eigenstates, we can
immediately write down the time development of the coherent state as 
\begin{equation}
\vert z,t)= 
{\Lambda _{0}(z)}\sum\limits_{n=0}^{+\infty }{Q_{n}(z)\,}\left[
\sum\limits_{\mu }{\sqrt{\Omega _{\mu }}p_{n}^{\ast }(E_{\mu })e^{-iE_{\mu
}t}{\left\vert E_{\mu }\right\rangle }+\int\limits_{\sigma _{c}}\sqrt{\Omega
(E)}p_{n}^{\ast }(E)e^{-iEt}{\left\vert E\right\rangle }dE}\right].
\label{GrindEQ__15_}
\end{equation}%
We will see in  section 5 an application of this relation to specific
systems.

\section{Special coherent states and their use a basis}

When we examine the general solution of the coherent states as given by the
infinite series expansion of Eq. \eqref{GrindEQ__10_} and %
\eqref{GrindEQ__12_}, we can identify some special states having interesting
properties. These are the ones where the parameter $z$ coincides with one of
the coefficients $\{ c_{n} \} _{n=0}^{\infty } .$ When $z=c_{\alpha } $, the
function $Q_{n} (c_{\alpha } )$ vanishes identically for all $n\ge \alpha +1$%
. If we define a matrix $\Lambda $ whose elements are $\Lambda _{n,\alpha }
=\Lambda _{n} (c_{\alpha } )\, $, then it is an upper triangular matrix. In
terms of this matrix, we can write the form of the special coherent state $%
\left|\left. c_{\alpha } \right)\right. $ simply as 
\begin{equation}  \label{GrindEQ__16_}
\left|\left. c_{\alpha } \right)\right. =\sum _{n=0}^{\alpha }{\left| \phi
_{n} \right\rangle} \, \Lambda _{n,\alpha }.
\end{equation}
Because of the finiteness of the sum, finding the state $\left|\left.
c_{\alpha } \right)\right. $ is grossly easier  than finding  $\left|\left.
z\right)\right. $ for general $z$. As examples, the first two of these
special states are $\left|\left. c_{0} \right)\right. ={\left| \phi _{0}
\right\rangle} $ and $\left|\left. c_{1} \right)\right. ={\left| \phi _{0}
\right\rangle} \Lambda _{0,1} +{\left| \phi _{1} \right\rangle} \Lambda
_{1,1} $. Additionally, it is clear that $\left|\left. c_{\alpha }
\right)\right. $ can be written as a linear combination of the first $%
(\alpha +1)$ members of the basis set. Conversely, it is then possible to
write each member of the basis ${\left| \phi _{n} \right\rangle} $ as a
linear combination of the first $(n+1)$ members of the special coherent
states. In fact, we can put the set $\{ \left|\left. c_{\alpha }
\right)\right. \} _{\alpha =0}^{\infty } $ into one-to-one correspondence
with the set $\{ {\left| \phi _{n} \right\rangle} \} _{n=0}^{\infty } $.
Therefore, similar to the basis, the set $\{ \left|\left. c_{\alpha }
\right)\right. \} _{\alpha =0}^{\infty } $ is complete. This is sufficient
condition to qualify this set to be designated as basis.  We
know that it is normalized by construction but not orthogonal.

The matrix $\Lambda $ allows us to construct $\left\vert \left. c_{\alpha
}\right) \right. $ from the set $\{{\left\vert \phi _{n}\right\rangle }%
\}_{n=0}^{\alpha }$ via Eq. \eqref{GrindEQ__16_}. For convenience, we need a
relation that enables us to construct ${\left\vert \phi _{n}\right\rangle }$
from the set $\{\left\vert \left. c_{\alpha }\right) \right. \}_{\alpha
=0}^{n}.$ For that purpose we use the inverse matrix $\bar{\Lambda}$ to the
matrix $\Lambda $. Similar to $\Lambda $, the matrix $\bar{\Lambda}$ is also
upper triangular satisfying 
\begin{equation}
\sum_{\alpha =0}^{m}\Lambda _{n,\alpha }\,\bar{\Lambda}_{\alpha ,m}=\delta
_{n,m}\quad ,\;\sum_{n=0}^{\beta }\bar{\Lambda}_{\alpha ,n}\,\Lambda
_{n,\beta }=\delta _{\alpha ,\beta }.  \label{GrindEQ__17_}
\end{equation}%
In order to move conveniently between the two bases, we use Eq. %
\eqref{GrindEQ__16_} and the above orthogonality relations to conclude that 
\begin{equation}
{\left\vert \phi _{n}\right\rangle }=\sum_{\alpha =0}^{n}\left\vert \left.
c_{\alpha }\right) \right. \,\bar{\Lambda}_{\alpha ,n}.  \label{GrindEQ__18_}
\end{equation}%
The expansion of a general coherent state $\left\vert \left. z\right)
\right. $ in terms of the basis $\{{\left\vert \phi _{n}\right\rangle }%
\}_{n=0}^{\infty }$, as in Eq. \eqref{GrindEQ__10_}, can now be written in
term of the basis $\{\left\vert \left. c_{\alpha }\right) \right. \}_{\alpha
=0}^{\infty }$ using the above relation as follows 
\begin{equation}
\left\vert \left. z\right) \right. =\sum_{n=0}^{\infty }\sum_{\alpha
=0}^{n}\left\vert \left. c_{\alpha }\right) \right. \,\bar{\Lambda}_{\alpha
,n}\Lambda _{n}(z).  \label{GrindEQ__19_}
\end{equation}%
We decouple the indices in the above double sums by using the relation \cite{McB}
\begin{equation}
\sum_{n=0}^{\infty }\sum_{\alpha =0}^{n}F(\alpha ,n-\alpha
)\,=\sum_{n=0}^{\infty }\sum_{\alpha =0}^{\infty }F(\alpha ,n)\,.
\label{GrindEQ__20_}
\end{equation}%
This leads to the following desired expansion in terms of the special basis, 
\begin{equation}
\left\vert \left. z\right) \right. =\sum_{\alpha =0}^{\infty }\left\vert
\left. c_{\alpha }\right) \right. \,K_{\alpha }(z)\quad ,\;K_{\alpha
}(z)=\sum_{n=0}^{\infty }\bar{\Lambda}_{\alpha ,n+\alpha }\,\Lambda
_{n+\alpha }(z)\,\,.  \label{GrindEQ__21_}
\end{equation}%
It is necessary to check that this formula reproduces the special coherent
states. For suppose that $z=c_{\beta }$, then $\Lambda _{n+\alpha
}(z)=\Lambda _{n+\alpha ,\beta }$ and 
\begin{equation}
K_{\alpha }(c_{\beta })=\sum_{n=0}^{\infty }\bar{\Lambda}_{\alpha ,n+\alpha
}\,\Lambda _{n+\alpha ,\beta }=\sum_{j=\alpha }^{\beta }\bar{\Lambda}%
_{\alpha ,j}\,\Lambda _{j,\beta }.
\end{equation}
The last sum vanishes identically for $\beta \leq \alpha -1$ , while for $%
\beta \geq \alpha $, $K_{\alpha }(c_{\beta })=\delta _{\alpha ,\beta }$ by the
orthogonality relation Eq. \eqref{GrindEQ__17_}. Hence $\left\vert \left.
z\right) \right. =\left\vert \left. c_{\beta }\right) \right. $. We now have
to find explicit the explicit values of $K_{\alpha }(z)$ in terms of the
basic parameters of the system. We first claim that the matrix elements of $%
\bar{\Lambda}\,$are given explicitly by 
\begin{equation}
\bar{\Lambda}_{\alpha ,m}=\left\vert 
\begin{array}{c}
{\;\quad \;\frac{1}{\Lambda _{\alpha ,\alpha }}\,\prod_{j=\alpha +1}^{m}(%
\frac{d_{j}}{c_{\alpha }-c_{j}})\quad \quad ,\,m\geq \alpha +1\quad } \\ 
{\frac{1}{\Lambda _{\alpha ,\alpha }}\quad \quad \quad \quad \quad \quad
\quad \quad \;,\,m=\alpha \,} \\ 
{\;\;\;\;0\quad \quad \quad \quad \quad \quad \quad \,\;\;\quad \quad
,\,m\leq \alpha -1}%
\end{array}%
\right. \,  \label{GrindEQ__22_}
\end{equation}%
We give the proof in Appendix A. We further state the following summation formula which will be 
helpful in finding the explicit form of the function $K_{\alpha }(z)$.

\begin{equation}
1+\sum_{n=1}^{\gamma }\prod_{j=\alpha }^{n+\alpha -1}(\frac{z-c_{j}}{%
c_{\alpha }-c_{j+1}})=\prod_{k=\alpha +1}^{\gamma +\alpha }(\frac{z-c_{k}}{%
c_{\alpha }-c_{k}})\quad ,\,\gamma \geq \alpha +1.  \label{GrindEQ__23_}
\end{equation}%
For the  proof of \eqref{GrindEQ__23_} see  Appendix B. Now we examine in detail the quantity $%
K_{\alpha }(z)$. 
\begin{eqnarray}
\,K_{\alpha }(z)&=&\sum_{n=0}^{\infty }\bar{\Lambda}_{\alpha ,n+\alpha
}\,\Lambda _{n+\alpha }(z) \cr
&=&\bar{\Lambda}_{\alpha ,\alpha }\,\Lambda _{\alpha }(z)\,\sum_{n=0}^{\infty
}\frac{\bar{\Lambda}_{\alpha ,n+\alpha }}{\bar{\Lambda}_{\alpha ,\alpha }}\,%
\frac{\Lambda _{n+\alpha }(z)}{\Lambda _{\alpha }(z)} \cr
&=&\bar{\Lambda}_{\alpha ,\alpha }\,\Lambda _{\alpha }(z)\,\sum_{n=0}^{\infty
}\prod_{j=\alpha +1}^{n+\alpha }\frac{d_{j}}{c_{\alpha }-c_{j}}%
\,\prod_{k=n}^{n+\alpha -1}\frac{z-c_{k}}{d_{k+1}} \cr
&=&\bar{\Lambda}_{\alpha ,\alpha }\,\Lambda _{\alpha }(z)\,\left[ {%
1+\,\sum_{n=1}^{\infty }\prod_{j=\alpha }^{n+\alpha -1}\frac{z-c_{j}}{%
c_{\alpha }-c_{j+1}}}\right]. 
\end{eqnarray}%
By Eq. \eqref{GrindEQ__23_}, the quantity in curly brackets is just $\prod_{k=\alpha
+1}^{\infty }(\frac{z-c_{k}}{c_{\alpha }-c_{k}})$. Thus, we have the major
result 
\begin{equation}
\,K_{\alpha }(z)=\bar{\Lambda}_{\alpha ,\alpha }\,\Lambda _{\alpha
}(z)\,\prod_{k=\alpha +1}^{\infty }(\frac{z-c_{k}}{c_{\alpha }-c_{k}}).
\end{equation}%
We now  use the facts that 
\begin{equation}
\,\bar{\Lambda}_{\alpha ,\alpha }=\frac{1}{\Lambda _{\alpha ,\alpha }}=\frac{%
1}{\Lambda _{0,\alpha }}\prod_{j=0}^{\alpha -1}(\frac{d_{j+1}}{c_{\alpha
}-c_{j}});\quad \Lambda _{\alpha }(z)=\Lambda _{0}(z)\prod_{j=0}^{\alpha -1}(\frac{%
z-c_{j}}{d_{j+1}}).
\end{equation}%
Thus, $\bar{\Lambda}_{\alpha ,\alpha }\,\Lambda _{\alpha }(z)=\frac{\Lambda
_{0}(z)}{\Lambda _{0,\alpha }}\prod_{j=0}^{\alpha -1}(\frac{z-c_{j}}{%
c_{\alpha }-c_{j}})$. Therefore, we  have another version for $K_{\alpha }(z)
$, namely 
\begin{equation}
K_{\alpha }(z)=\frac{\Lambda _{0}(z)}{\Lambda _{0,\alpha }}%
\prod_{j=0}^{\alpha -1}(\frac{z-c_{j}}{c_{\alpha }-c_{j}})\,\prod_{k=\alpha
+1}^{\infty }(\frac{z-c_{k}}{c_{\alpha }-c_{k}})=\frac{\Lambda _{0}(z)}{%
\Lambda _{0,\alpha }}\,\prod_{k\neq \alpha }^{\infty }(\frac{z-c_{k}}{%
c_{\alpha }-c_{k}}).  \label{GrindEQ__24_}
\end{equation}%
This leads to the following revealing form of the expansion of a general
coherent state $\left\vert \left. z\right) \right. $ in terms of the special
set of coherent states $\{\left\vert \left. c_{\alpha }\right) \right.
\}_{\alpha =0}^{\infty }$ , 
\begin{equation}
\frac{1 }{\Lambda _{0}(z)}\left\vert \left. z\right) \right.=\sum_{\alpha
=0}^{\infty}L_{\alpha }(z)\frac{ \;\left\vert \left. c_{\alpha }\right) \right.}{\Lambda
_{0,\alpha }},\quad L_{\alpha }(z)=\prod_{k\neq \alpha
}^{\infty }(\frac{z-c_{k}}{c_{\alpha }-c_{k}}).  \label{GrindEQ__25_}
\end{equation}%
This is just the realization of the Lagrange's interpolation scheme of the
set $\{\frac{\left\vert \left. c_{\alpha }\right) \right. }{\Lambda
_{0,\alpha }}\}_{\alpha =0}^{\infty }$ of values of the vector function $%
\frac{1 }{\Lambda _{0}(z)}\left\vert \left. z\right) \right.$ of the variable $%
z$ at the set of points $\{z=c_{\alpha }\}_{\alpha =0}^{\infty }$. We
recognize that the quantity $L_{\alpha }(z)$ is just the Lagrange kernel
\cite{WW}. From the rich literature on Lagrange interpolation scheme, we know
that there are other equivalent versions to the one depicted in Eq. %
\eqref{GrindEQ__25_}. We just quote them below without further
justification, 
\begin{equation}
L_{\alpha }(z)=\prod_{k\neq \alpha }^{\infty }(\frac{z-c_{k}}{c_{\alpha
}-c_{k}})\;;\;L(z)\equiv \prod_{\beta =0}^{\infty }(z-c_{\beta })\;;\;\sigma
_{\alpha }=\frac{1}{\prod_{\beta \neq \alpha }^{\infty }(c_{\alpha
}-c_{\beta })}
\end{equation}
\begin{equation*}
(i)\quad \frac{1}{\Lambda _{0}(z)}\left\vert \left. z\right) \right. 
=\sum_{\alpha =0}^{\infty }L_{\alpha }(z)\,\frac{ \left\vert \left. c_{\alpha
}\right) \right.}{\Lambda _{0,\alpha }}\;\quad ,\quad (ii)\quad \frac{1 }{\Lambda _{0}(z)}%
\left\vert \left. z\right) \right.=\,L(z)\,\sum_{\alpha
=0}^{\infty }\frac{\sigma _{\alpha }}{z-c_{\alpha }}\;\frac{\left\vert
\left. c_{\alpha }\right) \right. }{\Lambda _{0,\alpha }} 
\end{equation*}
\begin{equation}
(iii)\quad \frac{1}{\Lambda _{0}(z)}\left\vert \left. z\right) \right.=\,\,%
\frac{\sum_{\alpha =0}^{\infty }\frac{\sigma _{\alpha }}{z-c_{\alpha }}\;%
\frac{\left\vert \left. c_{\alpha }\right) \right. }{\Lambda _{0,\alpha }}}{%
\sum_{\alpha =0}^{\infty }\frac{\sigma _{\alpha }}{z-c_{\alpha }}\;}.\label{GrindEQ__26_}
\end{equation}

We must now show that form $(i)$, and equivalently the others, satisfies two
necessary properties. First, when $z=c_{\beta }$ it yields the coherent
state $\left\vert \left. c_{\beta }\right) \right. $. This easily follows
from the fact that $L_{\alpha }(c_{\beta })=\delta _{\alpha ,\beta }$.
Second, the form has to satisfy the defining equation for the coherent
state, namely, $A\left\vert \left. z\right) \right. =z\left\vert \left.
z\right) \right. $. This is demonstrated in Appendix C. Later in section 5,
we contrast these forms with the original representation of the coherent
state $\left\vert \left. z\right) \right. $in the basis $\{{\left\vert \phi
_{n}\right\rangle }\}_{n=0}^{\infty }$ by comparing the associated space
density $\rho (z;x)=\left\vert \left\langle x\right. \left\vert \left.
z\right) \right. \right\vert ^{2}$ of each form.

\section{\textbf{Examples}}

\subsection{The ground state as a coherent state.}

The ground state ${\left\vert E=0\right\rangle }$ is a coherent state $%
\left\vert \left. z\right) \right. $ with $z=0$. We illustrate this with two
examples.

\subsubsection{The Morse oscillator}

It is known that the one-dimensional Morse Hamiltonian 
\begin{equation}
-\frac{1}{2}\frac{d^{2}}{dx^{2}}+V_{0}(e^{-2\alpha x}-2e^{-\alpha x})+\frac{1%
}{2}\alpha ^{2}D^{2}
\end{equation}%
has a tridiagonal matrix representation in the complete orthonormal basis%
\begin{equation}\label{5.12}
\phi _{n}(x)=\sqrt{\frac{n!\,\alpha }{\Gamma (n+2\gamma +1)}}\,y^{\gamma +%
\frac{1}{2}}\,e^{-\frac{y}{2}}\,L_{n}^{(2\gamma )}(y)
\end{equation}
where the variable $y$ is $y(x)=\frac{\sqrt{8V_{0}}}{\alpha }\,e^{-\alpha x}$
, where $V_0$ and $\alpha$ are nonnegative given parameters of the oscillator, $D=\frac{\sqrt{2V_{0}}}{\alpha }-\frac{1}{2}$ and $\gamma $ is a free
scale parameter \cite{YM1}. Here , $L_n^{(\eta)}(\cdot)$ denotes the Laguerre polynomial (\cite{MOS}, p.239). The associated coefficients $\{c_{n},d_{n}\}_{n=0}^{%
\infty }$ with this Hamiltonian are given explicitly as%
\begin{equation}
c_{n}(\gamma )=\frac{\alpha }{\sqrt{2}}(n+\gamma +\frac{1}{2}%
-D)\;,\;d_{n}(\gamma )=\frac{-\alpha }{\sqrt{2}}\sqrt{n(n+2\gamma )}.
\end{equation}
Since $\gamma $ is a free scale parameter, and if the potential
supports one or more bound states, then we may choose $\gamma $ to have the
value $\gamma _{0}=D-\frac{1}{2}$. In that case  $c_0(\gamma_0)=0$ and hence	 the coherent state $\left.
\left\vert z=0\right. \right)=|c_0(\gamma_0)\rangle$. Therefore the ground state  simply has  one term in the expansion of Eq.\eqref{GrindEQ__10_}, namely, $\left\vert \left. z=0\right) \right. ={%
\left\vert \phi _{0}\right\rangle }$. In that case, we have by Eq.\eqref{5.12} 
\begin{equation}
\phi _{0}(x)=\sqrt{\frac{\alpha \,}{\Gamma (2D)}}\,y^{D}\,e^{-\frac{y}{2}}
\end{equation}%
which is indeed  the ground state wavefunction. It is interesting to note that if
there are more that one bound state, we can alternatively choose the free
scale parameter to have the value $\gamma _{0}=D-\frac{3}{2}$ . With this
choice, $c_{1}(\gamma _{1})=0$ and hence $\left\vert \left. z=0\right)
\right. =| c_{1}(\gamma _{1}))$. We know that%
\begin{equation}
\left\vert \left. c_{1}(\gamma _{1})\right) \right. ={\left\vert \phi
_{0}(\gamma _{1})\right\rangle }\Lambda _{0,1}(\gamma _{1})+{\left\vert \phi
_{1}(\gamma _{1})\right\rangle }\Lambda _{1,1}(\gamma _{1}).
\end{equation}
In this case, 
\begin{equation}
c_{0}(\gamma _{1})=\frac{-\alpha }{\sqrt{2}},\;d_{1}(\gamma_1)=\frac{-\alpha 
}{\sqrt{2}}\sqrt{(2D-2)}.
\end{equation}%
We also have%
\begin{equation}
Q_{1}(\gamma _{1})=\frac{c_{1}(\gamma _{1})-c_{0}(\gamma _{1})}{d_{1}(\gamma
_{1})}=\frac{-1}{\sqrt{(2D-2)}}.
\end{equation}
Thus, we have explicitly for the special coherent state,%
\begin{equation}
\left\vert \left. c_{1}(\gamma _{1})\right) \right. =\Lambda _{0,1}(\gamma
_{1})\left[\left\vert \phi _{0}(\gamma _{1})\right\rangle +\left\vert \phi
_{1}(\gamma _{1})\right\rangle Q_{1}(\gamma _{1})\right]
\end{equation}%
\begin{equation}
=\Lambda _{0,1}(\gamma _{1})\left[\left\vert \phi _{0}(\gamma
_{1})\right\rangle -\frac{1}{\sqrt{2D-2}}\left\vert \phi _{1}(\gamma
_{1})\right\rangle \right].
\end{equation}
If we now carefully substitute  the needed parameters in the
above equation, we find  again that the coherent state wavefunction 
\begin{equation}
\left\langle x\right. \left\vert \left. z=0\right) \right. =\left\langle
x\right. \left\vert \left. c_{1}(\gamma _{1})\right) \right. =\sqrt{\frac{%
\alpha \,}{\Gamma (2D)}}\,y^{D}\,e^{-\frac{y}{2}}
\end{equation}%
is the ground state.

\subsubsection{ The radial harmonic oscillator}

The $\ell -th$ partial wave harmonic oscillator Hamiltonian,%
\begin{equation}
-\frac{1}{2}\frac{d^{2}}{dr^{2}}+\frac{\ell (\ell +1)}{2r^{2}}+\frac{1}{2}%
\omega ^{2}r^{2}-(\ell +\frac{3}{2})\omega 
\end{equation}
has a tridiagonal matrix representation in the orthonormal basis 
\begin{equation}\label{eq5.12}
\phi _{n}(x)=\sqrt{\frac{n!\,(2\lambda )}{\Gamma (n+\nu +1)}}\,y^{\nu +\frac{%
1}{2}}\,e^{-\frac{y^{2}}{2}}\,L_{n}^{(\nu) }(y^{2})
\end{equation}%
where $y(r)=\lambda r,\;\nu =\ell +\frac{1}{2}$, and $\lambda $ is a free
scale parameter \cite{YM2}.  The Hamiltonian has a pure discrete spectrum with
energies $E_{n}=2n\omega $. The associated coefficients $\{c_{n},d_{n}%
\}_{n=0}^{\infty }$ with this Hamiltonian are given explicitly as 
\begin{equation}
c_{n}=\frac{(\lambda -\frac{\omega }{\lambda })}{\sqrt{2}}\sqrt{n+\ell +%
\frac{3}{2}}\;,\,d_{n}=\frac{(\lambda +\frac{\omega }{\lambda })}{\sqrt{2}}%
\sqrt{n}.
\end{equation}%
If we choose the free scale parameter to have the value $\lambda _{0}=\sqrt{%
\omega }$ , then all the $c_{n}$ vanish. Hence in the expansion %
\eqref{GrindEQ__10_} for the coherent state $\left. \left\vert z=0\right.
\right) $ only the first term survives. Therefore, $\left. \left\vert
z=0\right. \right) ={\left\vert \phi _{0}(\lambda _{0})\right\rangle }$. In
that case, we have from Eq. \eqref{eq5.12} 
\begin{equation}
\phi _{0}(r)=\sqrt{\frac{2\sqrt{\omega }\,}{\Gamma (\lambda +\frac{3}{2})}}%
\,(\sqrt{\omega }\,r)^{\ell +1}\,e^{-\frac{\omega r^{2}}{2}}\,
\label{GrindEQ__27_}
\end{equation}%
which is indeed the ground state wavefunction of the radial oscillator. It is curious to note that, because all of $c_{\alpha}(\lambda_0)$ vanish, all the special states $|c_{\alpha}(\lambda_0)\rangle$ become the ground state. This can also be confirmed by the fact that all the matrix elements  $\Lambda_{0,\alpha}$ for $\alpha\geq 1$ vanish because all of the functions $Q_n$ vanish for $n\geq 1$. What remains are  $\Lambda_{0,0}=1$ and $Q_0=1$, yielding $\langle r|c_{\alpha}(\lambda_0)\rangle=\phi_0(r)$, the ground state.  

\subsection{The coherent state of systems with known pure discrete energy
spectrum}

We consider the familiar case of the one-dimensional harmonic
oscillator with frequency $\omega $.\textit{\ }For this system $%
E_{n}=n\omega $, and hence $Q_{n}(z)=\frac{1}{\sqrt{n!}}\left(\frac{z}{\sqrt{\omega }}\right)^n
$. We then have $\Lambda _{0}(z)=e^{-\frac{\left\vert z\right\vert ^{2}}{%
2\omega }}$. Therefore, the coherent state function at time $t$ is given by 
\begin{eqnarray}
Z(x,t)\equiv \left\langle x\right. \left\vert \left. z,t\right) \right.&=&\Lambda _{0}(z)\sum_{n=0}^{\infty }Q_{n}(z)\,e^{-iE_{n}t}\,\phi _{n}(x)\cr
&=&\sqrt{\frac{\sqrt{\omega }}{\sqrt{\pi }}}\,e^{-\frac{\omega x^{2}}{2}%
}\,e^{-\frac{\left\vert z\right\vert ^{2}}{2\omega }}\,\sum_{n=0}^{\infty }%
\frac{(ze^{-i\omega t}/\sqrt{2\omega })^{n}}{n!}\,\,H_{n}(\sqrt{\omega }\,x),
\label{GrindEQ__28_}
\end{eqnarray}%
where $H_n(\cdot)$ denotes the Hermite polynomial (\cite{MOS}, p.249). Here, we have used the fact that the orthonormal energy eigenfunctions are the known%
\begin{equation}
\phi _{n}(x)=\sqrt{\frac{\sqrt{\omega }}{\sqrt{\pi }}}\,\frac{e^{-\frac{%
\omega x^{2}}{2}}}{\sqrt{2^{n}\,n!}}\,H_{n}(\sqrt{\omega }\,x).
\end{equation}
The sum in Eq.\eqref{GrindEQ__28_} is doable \cite{MOS}, and yields 
\begin{equation}
Z(x,t)\equiv \sqrt{\frac{\sqrt{\omega }}{\sqrt{\pi }}}\,e^{-\frac{\omega
x^{2}}{2}}\,e^{-\frac{\left\vert z\right\vert ^{2}}{2\omega }}\,\exp \left( 2%
\sqrt{\omega }\,x\right(\frac{ze^{-i\omega t}}{\sqrt{2\omega }}\left)-\,\right(\frac{%
ze^{-i\omega t}}{\sqrt{2\omega }}\left)^{2}\,\right) . \label{GrindEQ__29_}
\end{equation}
If we now write $z=\left|z\right|e^{i\varphi } $, and define the density of
the coherent state $\rho (z;x,t)\equiv \, \left|Z(x,t)\right|^{2} $, we finally have 
\begin{equation}  \label{GrindEQ__30_}
\rho (z;x,t)\equiv \sqrt{\frac{\omega }{\pi } } \, \, \exp\left(-\omega (x-\frac{%
\sqrt{2} \, \left|z\right|\, }{\omega } \cos (\omega t-\varphi )\, )^{2} \right) \,.
\end{equation}
This is just a Gaussian density function of constant width and with center $%
\bar{x}$ oscillating harmonically as 
\begin{equation}  \label{GrindEQ__31_}
\bar{x}(z,t)=\, \frac{\sqrt{2} \, \, }{\omega }\left|z\right| \cos (\omega
t-\varphi ).
\end{equation}
\subsection{Time development of special coherent state}

Even with the availability of Eq. \eqref{GrindEQ__15_} to describe the time
development of a general coherent state, it is a considerable challenge to
find the answer in close form. However, the difficulty is diminished
substantially if we able to utilize the special coherent state to act in
place of the general coherent state. For, if the basis is endowed with a
free scale parameter $\lambda $, it may be possible to choose $\lambda $ to
have the value $\lambda _{\alpha } $ such that $z=c_{\alpha } (\lambda
_{\alpha } )$. In that case, $\left|\left. z\right)\right. =\left|\left.
c_{\alpha } (\lambda _{\alpha } )\right)\right. $ and the calculation
difficulty is considerably reduced. We illustrate this case, by examining
the time development of coherent state associated with radial harmonic
oscillator and another associated with a radial free particle.
\subsubsection{Time development of the coherent state of the radial harmonic
oscillator}

For this system, $c_{n}(\lambda )=\frac{(\lambda -\frac{\omega }{\lambda })}{%
\sqrt{2}}\sqrt{n+\ell +\frac{3}{2}}$. Suppose the value of $z$ is compatible
with the restriction on $\lambda $ (which is $Re(\lambda )> 0$) so
that we can choose $\lambda $ to have the value%
\begin{equation}
\lambda =\lambda _{0}\equiv \frac{z}{\sqrt{2\ell +3}}+\sqrt{\frac{z^{2}}{%
2\ell +3}+\omega }.
\end{equation}
This immediately makes $\left\vert \left. z\right) \right. =\left\vert
\left. c_{0}(\lambda _{0})\right) \right. $. This, in turns,  means that $%
\left\vert \left. z\right) \right. =\,{\left\vert \phi _{0}(\lambda
_{0})\right\rangle }$. From Eq. \eqref{GrindEQ__15_}, and since the spectrum
is purely discrete, we have 
\begin{equation}
\,\left\vert \left. z,t\right) \right. =\sum_{\mu =0}^{\infty }\sqrt{\Omega
_{\mu }(\lambda _{0})}p_{n}^{\ast }(E_{\mu },\lambda _{0})e^{-iE_{\mu }t}{%
\left\vert E_{\mu }\right\rangle }  \label{GrindEQ__32_}
\end{equation}%
where we have shown the explicit the dependence on $\lambda _{0}(z)$.
Noting that the discrete density is \cite{YM2}
\begin{equation}
\Omega _{\mu }(\lambda _{0})=\frac{\Gamma (\mu +\nu +1)}{\mu !\,\,\Gamma
(\mu +\nu +1)}\,\tau ^{\mu }\,(1-\tau )^{\nu +1}\;,\,\tau =(\frac{\lambda
_{0}^{2}-\omega }{\lambda _{0}^{2}+\omega })^{2},
\end{equation}%
we have 
\begin{equation}
\psi (z;r,t)=\left\langle r\right. \,\left\vert \left. z,t\right) \right. =
\left( \frac{2\sqrt{\omega }}{\Gamma (\nu +1)}\right) ^{\frac{1}{2}}{\,\frac{%
[1-\tau ]^{(1+\nu )/2}}{[1-y(t)]^{(1+\nu )}}\,[\sqrt{\omega }\,r]^{(\ell
+1)}\,e^{-\frac{\omega r^{2}}{2}\frac{1+y(t)}{1-y(t)}},}
\end{equation}
\begin{equation}
y(t)=\sqrt{\tau }\,e^{-2i\omega t},\,\nu =\ell +\frac{1}{2}.\label{GrindEQ__33_}
\end{equation}
As a necessary quality check on this result, we note that $\psi (z;r,0)=\phi
_{0}(r,\lambda _{0})$. The space density of coherent state is just $\rho
(z;r,t)=\left\vert \psi (z;r,t)\right\vert ^{2}$, 
\begin{equation}
{\rho (z;r,t)=\left\vert \psi (z;r,t)\right\vert ^{2}=\,} {\frac{2[\omega G(t)]^{\nu +1}}{\Gamma (\nu +1)}\,r^{2\ell +2}e^{-\omega
r^{2}G(t)}\;,\,G(t)=\frac{1-\tau }{1+\tau -2\sqrt{\tau }\,\cos (2\omega t)}}.
\label{GrindEQ__34_}
\end{equation}%
The average position $\bar{r}(t)$ of the coherent state is computed via the
relation%
\begin{equation}
\bar{r}(t)=\int_{0}^{\infty }r\rho (z;r,t)\,dr.
\end{equation}%
The result is 
\begin{equation}
\bar{r}(t)=\frac{\Gamma (\nu +\frac{3}{2})}{\Gamma (\nu +1)}\sqrt{\frac{%
\lambda _{0}^{4}\,\sin ^{2}(\omega t)+\omega ^{2}\,\cos ^{2}(\omega t)}{%
\omega ^{2}\lambda _{0}^{2}}}.
\end{equation}%
We show in Fig.1 the sinusoidal nature of the time
development of $\bar{r}(t)$ and in Fig.2  a phase diagram
of $\frac{d}{dt}\bar{r}(t)$ versus $\bar{r}(t)$.
\subsubsection{Time development of the coherent state of the radial free
particle}

In contrast to the above example of the radial harmonic oscillator where the
energy spectrum is purely discrete, we work out the same problem but for a
radial free particle whose energy spectrum is purely continuous. The
Hamiltonian associated with this problem is tridiagonal in the same
orthonormal basis as the previous one, but now yielding the quantity $%
c_{n}(\lambda )=\frac{\lambda }{\sqrt{2}}\sqrt{n+\ell +\frac{3}{2}}$. If we
now choose the free scale parameter $\lambda $ to have the value $\lambda
_{0}=\frac{2z}{\sqrt{2\ell +3}}$ , then $\left\vert \left. z\right) \right.
=\left\vert \left. c_{0}(\lambda _{0})\right) \right. $. From Eq. %
\eqref{GrindEQ__15_}, we have 
\begin{equation}
\,\left\vert \left. z,t\right) \right. =\,\int_{0}^{\infty }\sqrt{\Omega (E)}%
\,e^{-iEt}{\left\vert E\right\rangle }dE.  \label{GrindEQ__35_}
\end{equation}%
Noting that the continuous density is \cite{YM2}: 
\begin{equation}
\Omega (E,\lambda _{0})=\frac{2}{\lambda _{0}^{2}\,\,\Gamma (\nu +1)}\,(%
\frac{2E}{\lambda _{0}^{2}})^{\nu }\,e^{-\frac{2E}{\lambda _{0}^{2}}}
\end{equation}
and $\langle r|E \rangle=\sqrt{r}J_{\nu} (kr)$ with $\frac{k^2}{r}=E$, we get 
\begin{equation}
\psi (z;r;t)=\,[\frac{\beta (t)}{\lambda }]^{\nu +1}\sqrt{\frac{2\beta (t)}{%
\Gamma (\nu +1)}}\;[\beta (t)\;r]^{\ell +1}\;e^{-\frac{[\beta (t)\;r]^{2}}{2}%
}\;,\;\beta (t)=\frac{\lambda }{\sqrt{1+i\lambda ^{2}t}}.
\label{GrindEQ__36_}
\end{equation}%
The space density of the coherent state becomes 
\begin{equation}
\rho (r,t)=\frac{2}{\Gamma (\nu +1)}\left( \frac{\beta \beta ^{\ast }}{%
\lambda }\right) ^{2\nu +2}r^{2\ell +2}\;e^{-\left( \frac{\beta ^{2}+\beta
^{\ast 2}}{2}\right) \,\,r^{2}}.  \label{GrindEQ__37_}
\end{equation}%
This translates into 
\begin{equation}
\rho (z;r,t)=\frac{2}{\Gamma (\nu +1)}\frac{\lambda _{0}}{(1+\lambda
_{0}^{4}t^{2})^{\nu +1}}(\lambda _{0}r)^{2\ell +2}\;e^{-\,\,\frac{\lambda
^{2}r^{2}}{1+\lambda _{0}^{4}t^{2}}}.  \label{GrindEQ__38_}
\end{equation}%
The average position of the coherent state is%
\begin{equation}
\bar{r}(t)=\frac{\Gamma (\nu +\frac{3}{2})}{\Gamma (\nu +1)}\sqrt{\frac{%
1+\lambda _{0}^{4}\,t^{2}}{\lambda _{0}^{2}}}.
\end{equation}
Asymptotically, for large time, the average position progresses linearly
with time,%
\begin{equation}
{\mathop{\lim }\limits_{t\rightarrow \infty }}\bar{r}(t)=\frac{\Gamma (\nu +%
\frac{3}{2})}{\Gamma (\nu +1)}\,\lambda _{0}t
\end{equation}
as expected. We show in Fig.3 and Fig.4 the
time dependence of $\bar{r}(t)$ and $\frac{d}{dt}\bar{r}(t)$ respectively.
We note in passing that the results of the previous example on the radial
harmonic oscillator reduce to the above results in the limit of vanishing
frequency $\omega $.
\subsection{Performance of the Lagrange form of the coherent states}

Here we consider, as a reference, the coherent state $(z=0.83)$ of the Morse
oscillator as represented by a converged finite sum of order $N$ of Eq. \eqref{GrindEQ__10_}. We also construct a converged finite sum of
the three Lagrange forms of the coherent state. For each, we find the real
space density and compare its graph with that of the reference case in Fig.5. We note that all methods give practically identical
results. This is not surprising since for the same scale parameter $\gamma$ 
 and order of approximation $N$, the finite basis $\{ {\left| \phi _{n}
\right\rangle} \} _{n=0}^{N-1} $ span the same subspace as that spanned by
the finite basis $\{ \left|\left. c_{\alpha } \right)\right. \} _{\alpha
=0}^{N-1} \; $. Had we chosen to find a scale parameter $\gamma_0$ such that $z=c_0(\gamma_0)$, the coherent state can also be represented as $|c_0(\gamma_0))$. In that case $\gamma_0=4.646$. The reader can easily verify that any of the four densities displayed in Fig.5 is practically identical to the density $|\langle x|c_0(\gamma_0) )|^2$. 

\section{Comments }

The over-completeness of the coherent states is usually handled through a
demonstration of the resolution of the identity as in Eq. \eqref{eq1.1}. In this paper, we tackle
this issue by exhibiting a proper subset of coherent states, namely, $%
\{\left\vert \left. c_{\alpha }\right) \right. \}_{\alpha =0}^{\infty }\;$,
which is complete.

Also, it is known that the canonical coherent state $\{ \left|\left.
c_{\alpha } \right)\right. \} _{\alpha =0}^{\infty } \; $can be generated
from the fiducial state $\left|\left. 0\right)\right. $, that satisfies $%
A\left|\left. 0\right)\right. =0$, through the relation $\left|\left.
z\right)=\right. \, e^{(zA^{\dag } -z^{*} A)} \left|\left. 0\right)\right. $%
. In our approach, because the forward shift and back ward shift operators, $%
A$ and $A^{\dag } $ , are not true lowering and raising operators, the terms 
$(A^{\dag } )^{n} \left|\left. 0\right)\right. $exhibit no particular
simplicity. Instead, we use the general basis $\{ {\left|
\phi _{n} \right\rangle} \} _{n=0}^{\infty } $ which tridiagonalizes the
Hamiltonian. Often times, this basis allows the freedom of choosing a scale
parameter flexibly. This is a great advantage which we have shown how to exploit
effectively. Furthermore, we introduced the complete subset of
coherent states, $\{ \left|\left. c_{\alpha } \right)\right. \} _{\alpha
=0}^{\infty } \; $to be used as an indigenous basis to describe any general
coherent state $\left|\left. z\right)\right. $. In this basis $\left|\left.
z\right)\right. $ is really just  the result of interpolating ``\textit{\`a la
Lagrange}" the special coherent states $\{ | c _{\alpha} )\} _{\alpha=0}^{\infty } $. 

If the Hamiltonian enjoys additional symmetries, such as shape-invariance \cite{YM3}, it is then expected that this will be reflected in additional computational convenience of the associated coherent states.   \bigskip\\
\textbf{Appendix A}

We show here that the matrix $\bar{\Lambda }$ whose elements are given by Eq. %
\eqref{GrindEQ__22_}, is actually the inverse of the matrix $\Lambda $. We
first show that $\bar{\Lambda }\, \Lambda =1$. Because both matrices are
upper triangular, we have
\begin{equation}
(\bar{\Lambda }\, \Lambda )_{\alpha ,\beta } =\sum _{n=0}^{\infty }\bar{%
\Lambda }_{\alpha ,n} \, \Lambda _{n,\beta } =\sum _{n=\alpha }^{\beta }\bar{%
\Lambda }_{\alpha ,n} \, \Lambda _{n,\beta }. \tag{A.1}
\end{equation}
Since the matrix $\bar{\Lambda }\, \Lambda $ is also upper triangular, the
element $(\bar{\Lambda }\, \Lambda )_{\alpha ,\beta } $ must vanish for $%
\alpha \ge \beta +1$. If $\alpha =\beta $, then $(\bar{\Lambda }\, \Lambda
)_{\beta ,\beta } =\, \bar{\Lambda }_{\beta ,\beta } \Lambda _{\beta ,\beta
} =1$. Finally, if $\beta \ge \alpha +1$, we have
\begin{equation}
{(\bar{\Lambda }\, \Lambda )_{\alpha ,\beta } =\, \sum _{n=\alpha }^{\beta }%
\bar{\Lambda }\, _{\alpha ,n} \, \, \Lambda _{n,\beta } } 
{=\bar{\Lambda }\, _{\alpha ,\alpha } \Lambda _{\alpha ,\beta } +\bar{%
\Lambda }\, _{\alpha ,\alpha +1} \Lambda _{\alpha +1,\beta } +\ldots+}\bar{%
\Lambda }\, _{\alpha ,\beta -1} \Lambda _{\beta -1,\beta } +\bar{\Lambda }\,
_{\alpha ,\beta } \Lambda _{\beta ,\beta }. \tag{A.2}
\end{equation}
This means that
\begin{eqnarray*}
(\bar{\Lambda }\, \Lambda )_{\alpha ,\beta } &=&\, \bar{\Lambda }\, _{\alpha
,\alpha } \Lambda _{\alpha ,\beta } \left[ 1+\frac{\bar{\Lambda }\, _{\alpha
,\alpha +1} \Lambda _{\alpha +1,\beta } }{\bar{\Lambda }\, _{\alpha ,\alpha
} \Lambda _{\alpha ,\beta } }+\ldots+\frac{\bar{%
\Lambda }\, _{\alpha ,\beta -1} \Lambda _{\beta -1,\beta } }{\bar{\Lambda }%
\, _{\alpha ,\alpha } \Lambda _{\alpha ,\beta } } +\frac{\bar{\Lambda }\,
_{\alpha ,\beta } \Lambda _{\beta ,\beta } }{\bar{\Lambda }\, _{\alpha
,\alpha } \Lambda _{\alpha ,\beta } } \right]\cr  
&=&\bar{\Lambda }\, _{\alpha ,\alpha } \Lambda _{\alpha ,\beta } \left[ 1+\sum
_{n=\alpha +1}^{\beta }\prod _{j=\alpha +1}^{n}(\frac{c_{\beta } -c_{j-1} }{%
c_{\alpha } -c_{j} } ) \right]
\end{eqnarray*}
\begin{equation}
\hspace*{-3.2cm}=\bar{\Lambda }\, _{\alpha ,\alpha } \Lambda _{\alpha ,\beta } \left[ 1+\sum
_{n=1}^{\beta -\alpha }\prod _{j=\alpha }^{n+\alpha -1}(\frac{c_{\beta }
-c_{j} }{c_{\alpha } -c_{j+1} } ) \right]	.\tag{A.3}
\end{equation}
By appealing to Eq. \eqref{GrindEQ__23_} while putting $\gamma
=\beta -\alpha $ and substituting $z=c_{\beta } $, we see that the last
expression vanishes.\medskip\\
\textbf{Appendix B}

With $ \gamma \ge \alpha +1$, we verify that the left-hand side (LHS)
of Eq. \eqref{GrindEQ__23_} of the Lemma is the same as the right-hand side
(RHS).
\begin{eqnarray*}
LHS&=&1+(\frac{z-c_{\alpha } }{c_{\alpha } -c_{\alpha +1} } )+(\frac{%
z-c_{\alpha } }{c_{\alpha } -c_{\alpha +1} } )(\frac{z-c_{\alpha +1} }{%
c_{\alpha } -c_{\alpha +2} } )+...+\prod _{j=\alpha }^{\gamma +\alpha -1}(%
\frac{z-c_{j} }{c_{\alpha } -c_{j+1} } ) \cr 
&=&(\frac{z-c_{\alpha +1} }{c_{\alpha } -c_{\alpha +1} } )+(\frac{z-c_{\alpha
} }{c_{\alpha } -c_{\alpha +1} } )(\frac{z-c_{\alpha +1} }{c_{\alpha }
-c_{\alpha +2} } )+...+\prod _{j=\alpha }^{\gamma +\alpha -1}(\frac{z-c_{j} 
}{c_{\alpha } -c_{j+1} } ) \cr
&=&(\frac{z-c_{\alpha +1} }{c_{\alpha } -c_{\alpha +1} } )\left[ 1+(\frac{%
z-c_{\alpha } }{c_{\alpha } -c_{\alpha +2} } )+...+(\frac{z-c_{\alpha } }{%
c_{\alpha } -c_{\alpha +2} } )\prod _{j=\alpha +2}^{\gamma +\alpha -1}(\frac{%
z-c_{j} }{c_{\alpha } -c_{j+1} } ) \right] =...
\end{eqnarray*}
\begin{equation}
=(\frac{z-c_{\alpha +1} }{c_{\alpha } -c_{\alpha +1} } )(\frac{z-c_{\alpha
+2} }{c_{\alpha } -c_{\alpha +2} } )\ldots\left[ 1+(\frac{z-c_{\alpha } }{c_{\alpha
} -c_{\gamma +\alpha } } )\right] =\prod _{j=\alpha +1}^{\gamma +\alpha }(\frac{%
z-c_{j} }{c_{\alpha } -c_{j} } ) =RHS.
\tag{B.1}
\end{equation}
\bigskip\\
\textbf{Appendix C}

Here, we demonstrate that the Lagrange expansion of a coherent state
satisfies the defining equation for the coherent state, namely, $%
A\left|\left. z\right)\right. =z\left|\left. z\right)\right. $. The key to
the proof is that if a function $f(z)$ is any polynomial in $z$ of order $n$, then the
Lagrange interpolation $f(z)=\sum _{\alpha =0}^{n}f(c_{\alpha } ) \;
L_{\alpha } (z)$ is exact. Therefore,
\begin{eqnarray*}
A\left|\left. z\right)\right. &=&\Lambda _{0} (z)\sum _{\alpha =0}^{\infty }%
\frac{c_{\alpha } \left|\left. c_{\alpha } \right)\right. }{\Lambda
_{0,\alpha } } \; L_{\alpha } (z)\cr
&=&\Lambda _{0} (z)\sum _{\alpha =0}^{\infty
}\sum _{n=0}^{\alpha }c_{\alpha } \, L_{\alpha } (z) \frac{\Lambda
_{n,\alpha } }{\Lambda _{0,\alpha } } \; {\left| \phi _{n} \right\rangle} \cr
&=&\Lambda _{0} (z)\sum _{\alpha =0}^{\infty }\sum _{n=0}^{\infty
}c_{n+\alpha } \, L_{n+\alpha } (z) \frac{\Lambda _{n,n+\alpha } }{\Lambda
_{0,n+\alpha } } \; {\left| \phi _{n} \right\rangle}
\end{eqnarray*}
\begin{equation}
 \hspace*{1cm}=\Lambda _{0} (z)\sum
_{n=0}^{\infty }{\left| \phi _{n} \right\rangle} \sum _{\alpha =0}^{\infty
}c_{n+\alpha } \, L_{n+\alpha } (z) \frac{\Lambda _{n,n+\alpha } }{\Lambda
_{0,n+\alpha } } . 
\tag{C.1}
\end{equation}
We decoupled the connection between the two indices according to the
rule of Eq. \eqref{GrindEQ__20_}. Here, we have extended the summation  by
adding terms with zero contribution. Now,
\begin{eqnarray*}
A\left|\left. z\right)\right. &=&\Lambda _{0} (z)\sum _{n=0}^{\infty }{\left|
\phi _{n} \right\rangle} \sum _{k=n}^{\infty }c_{k} \, L_{k} (z) \frac{%
\Lambda _{n,k} }{\Lambda _{0,k} }\cr
& =&\Lambda _{0} (z)\sum _{n=0}^{\infty }{%
\left| \phi _{n} \right\rangle} \sum _{k=0}^{\infty }c_{k} \, L_{k} (z) 
\frac{\Lambda _{n,k} }{\Lambda _{0,k} } \; 
\end{eqnarray*}
\begin{equation}
\hspace*{2cm}=\Lambda _{0} (z)\sum _{n=0}^{\infty }{\left| \phi _{n} \right\rangle} \sum
_{k=0}^{\infty }c_{k} \, L_{k} (z) \prod _{j=0}^{n-1}(\frac{c_{k} -c_{j} }{%
d_{j+1} } ). \tag{C.2}
\end{equation}
The inner summation is the Lagrange interpolation of polynomial functions.
Hence,
\begin{equation}
A\left|\left. z\right)\right. =\; \Lambda _{0} (z)\sum _{n=0}^{\infty }{%
\left| \phi _{n} \right\rangle} \, z\prod _{j=0}^{n-1}(\frac{z-c_{j} }{%
d_{j+1} } ) =z\sum _{n=0}^{\infty }{\left| \phi _{n} \right\rangle} \,\Lambda _{n} (z) =z\, \left|\left. z\right)\right.. \tag{C.3}
\end{equation}
\\
\textbf{Data availability}
\medskip\\
The data that support the findings of this study are available within the
article.
\bigskip\\
\textbf{FIGURE CAPTIONS:}

Fig. 1: The sinusoidal behavior of the average position of the coherent
state of a radial harmonic oscillator with frequency $\omega =2$, angular
momentum $L=1$ and $z=3.$ The value of the  spacial scale parameter is $\gamma_0=3.291$

Fig. 2: The average position vs its time development of the coherent state
of a radial harmonic oscillator with frequency $\omega =2$, angular momentum 
$L=1$ and $z=3.$

Fig. 3: The average position of the coherent state of a radial free particle
with angular momentum $L=1$ and $z=3.$ The value of the special scale
parameter is $\gamma _{0} =2.863$.

Fig. 4: The time development of the velocity of the average position of  the coherent
state of a radial free particle with angular momentum $L=1$ and $z=3.$ 

Fig. 5: The coherent state space density for Morse oscillator with $V_{0}
=10 $ and $\alpha =1$ which supports four bound states. The graphs show the
comparison of the various methods with the reference case $\varrho \psi=|\langle x|z)|^2$ with $|z)$ resulting from the
use of the expansion of Eq.\eqref{GrindEQ__10_} with $\gamma =3$ and order
of approximation $N=10$. The graphs $\varrho_I(x),\:\varrho_{II}(x)$ and $\varrho_{III}(x)$ are for the three forms of the Lagrange interpolation schemas $i,\;ii$ and $iii$ respectively in Eq. \eqref{GrindEQ__26_}, for the  same choices of scale parameter and order of approximation.

\begin{figure}[H]
\centering
\includegraphics[scale=1]{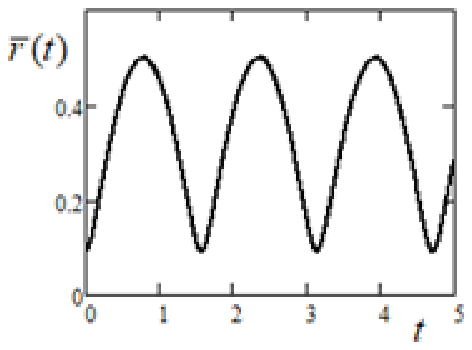}
\caption{}
\end{figure}

\begin{figure}[H]
\centering
\includegraphics*[scale=1]{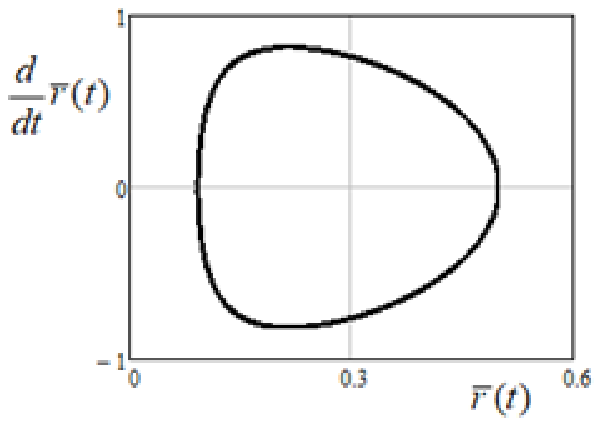} 
\vspace*{-1cm}
\caption{}
\end{figure}

\begin{figure}[H]
\centering
\includegraphics[scale=1]{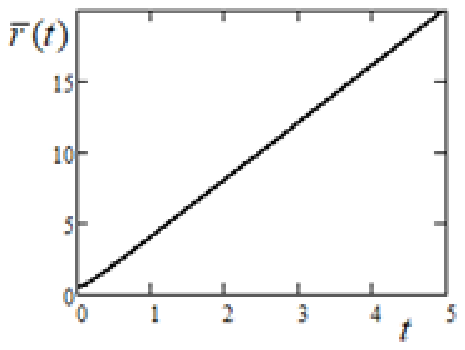} 
\vspace*{-1cm}
\caption{}
\end{figure}

\begin{figure}[H]
\centering
\includegraphics[scale=1]{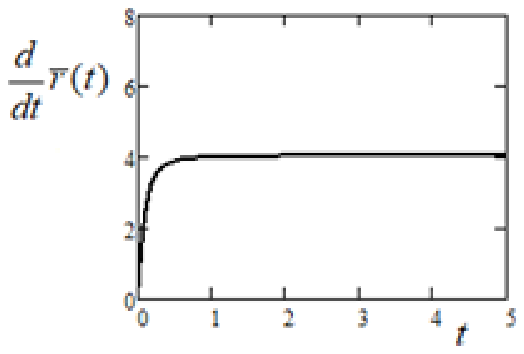} 
\vspace*{-1cm}
\caption{}
\end{figure}

\begin{figure}[H]
\centering
\includegraphics[scale=1]{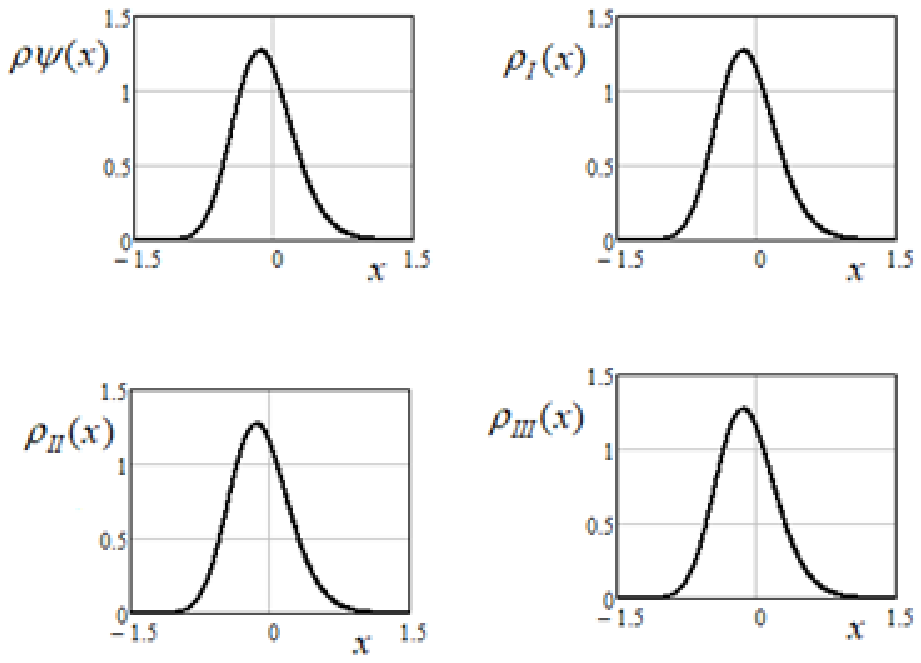} 
\caption{}
\end{figure}

\footnotesize{
 \bibliographystyle{plain}

}

\end{document}